\documentstyle[12pt]{article}
\newcommand{\beq}{\begin{equation}}
\newcommand{\eeq}{\end{equation}}
\newcommand{\beqa}{\begin{eqnarray}}
\newcommand{\eeqa}{\end{eqnarray}}
\newcommand{\beqar}{\begin{eqnarray*}}
\newcommand{\eeqar}{\end{eqnarray*}}
\newcommand{\tr}{{\rm tr}}

\def \vr {\hat\varrho}

\def \prob {{\cal P}rob}
\def \la {\langle}
\def \ra {\rangle}
\def \up {\uparrow}
\def \down {\downarrow}
\def \o {{\cal O}}
\def \h {{\cal H}}
\def \s {{\cal S}}

\def \e {{\cal E}}

\begin{document}
\begin{titlepage}
\thispagestyle{empty}
\begin{flushright}
TP-95-001\\
January 1995
\end{flushright}
\begin{center}
{\bf \Large Interaction with a pre and post selected environment \\
and recoherence}\\
\vspace{.4in}
B. Reznik
\footnote{\it e-mail: reznik@physics.ubc.ca}\\
\medskip
{ \small\it  Department of Physics}\\
{\small \it University of British Columbia}\\
{\small\it 6224 Agricultural Rd. Vancouver, B.C., Canada V6T 1Z1}\\
\end{center}
\vspace{.5in}
\begin{center}
\begin{minipage}{5in}
\begin{center}
{\large\bf Abstract}
\end{center}
{\small
The interaction of an open system $\s$ with a pre- and post-selected
environment
is studied.
In general, under such circumstances  $\s$ can not  be described in terms of a
density matrix, {\it even when $\s$ in not post-selected}. However, a simple
description in terms of a two-state (TS) is always available.  The two-state of
$\s$ evolves in ti
me from an initially  `pure' TS to a `mixed' TS and  back to a final `pure' TS.
This generic process is governed by a modified Liouville equation,  which is
derived.
For a  sub-class of observables, which can still be described by an ordinary
density matrix, this  evolution generates  recoherence to a final pure state.
In some cases post-selection  can even suppress  any decoherence.
}
\end{minipage}
\end{center}
\end{titlepage}

\section{Introduction}

The  interaction  of an  open quantum system with an environment \cite{openrev}
 is traditionally analyzed while assuming a given, not necessarily known,
initial state of the total closed system.
In this case the open system can be described by a reduced density matrix which
is obtained by tracing over the unknown environment's degrees of freedom.
In this work we investigate circumstances in which the environment, and
possibly also the open system, are bound to satisfy, not only an initial
condition, but also a second final condition. In other words, we shall consider
the interaction of a pre- and
post selected environment with an open system.\cite{protective}
Although, under usual circumstances, such a post selection is not realized,
it is in principle not forbidden.  Quantum Mechanics is (dynamically) time
symmetric, and it is possible to conceive situations in which the initial and
final conditions
are selected according to some `dynamical principle' (e.g. \cite{hawhar}).

We shall show that  when the environment is post selected,  the system can not
generally be described  in terms of a reduced density matrix. At  any
intermediate time, between the pre- and post-selection, there exists no pure or
mixed state, which yields
 the correct probabilities for measurements
in the open system, {\it even when the open system is not post-selected}.
We suggest that is such cases, it is preferable, both practically and
conceptually,  to describe the open system by a new object which is a
generalization of the density matrix.

It was recently suggested, that  a quantum system
should basically be described by an extension of the ordinary quantum state (or
density matrix) called a ``two-state'' (TS), which is determined by two,
initial and final, conditions \cite{vaidman,symm}. In the following we apply
the formalism developed i
n Ref. \cite{symm} to this problem. The probabilities for any measurement in
the open system are shown to be derived from a reduced TS, i.e. the TS obtained
by tracing over the environment's degrees of freedom.
When the initial and final state of the environment are given by pure states,
this reduced TS evolves in time from an initial `pure TS' to a
a `mixed TS' (of entangled form) at intermediate times, and finally back to a
pure TS. Therefore,  the effect of post-selecting the environment is to
``recohere  the TS''. This process is dynamically expressed by a modified
Liouville equation. As we shall
 show, the coefficients of the new terms in the this equation are time
dependent, and tuned in such a way that the TS finally ``recoheres''.

It is well known,  that interaction with an environment often causes
decoherence in the open system. (For example see:
\cite{sen,davydov,zurek,joos,unruh}).
In our case of post-selection, although the description in terms of a (pure or
mixed) density matrix is generally invalidated,  one can still find
 an effective  density matrix for a limited class of observables.
We show that the post-selection causes this effective density matrix to
recohere to a final pure state \cite{re-coherence}.
In some cases, depending on the nature of the interaction, post-selection can
suppress any decoherence.

This article proceeds as follows. In the next section we review shortly the
two-state formalism of quantum mechanics and elaborate on some relevant
details.
In Section 3. we apply this formalism to the case of a pre and post selected
system. A simple solvable example is given in Section 4.
The modified Liouville equation which is satisfied by the two-state is derived
in the last section using a perturbative approximation scheme and is applied to
some cases.
In the following we set $\hbar=1$.

\section{Quantum mechanics in terms of two-states}

Two-states are  particularly suitable in situations with two or more condition
on a single quantum system.
We now briefly review this formalism following Reference \cite{symm}, and
elaborate further on some relevant issues.

Consider a  system $\s$  with a given Hamiltonian $H_s$. Let us assume that
at $t_1$ and $t_2$ a complete set of measurements determine the states of $\s$
to be $|\psi_{in}(t_1)\ra$ and $|\psi_{out}(t_2)\ra$, respectively.
Now consider  an ensemble of such identical systems which is defined by the
latter two conditions. We are interested in probability distributions of
observables that are  measured in some intermediate time $t_2>t>t_1$. The
peculiarity of such a situation
is that in general (as we shall see) these probabilities can not be derived
from a single wave function or density matrix. It was therefore, suggested that
the ``state'' of $\s$ at intermediate times should be described by a
generalization of the ordinary
 wave function, which we call a `two-state'. Generically, a TS, which we denote
by $\vr$, is a non-Hermitian operator with the form:
\beq
\vr \ = \  |...\ra \la ...|
\label{generic}
\eeq
At the left and right slots of $\vr$  one inserts the information due to the
conditions at the $t_1$ and $t_2$ respectively.
In the case of a closed system $\s$ we have:
\beq
\vr(t) = U(t-t_1)|\psi_{in}\ra\la\psi_{out}|U^\dagger(t_2-t)=
|\psi_{in}(t)\ra\la\psi_{out}(t)|
\eeq
where $U(t)$ is the unitary evolution operator.

More generally, two-states are elements of a Hilbert space $\h_{II}$,  which is
defined as follows.
Given by a Hilbert  space of states ${\cal H}_I=\lbrace|\alpha\rangle\rbrace$,
we can construct the linear  space ${\cal
H}_{II}=\lbrace|\alpha\rangle\langle\beta|\rbrace$,
where $|\alpha\rangle$ and $|\beta\rangle$
are any two elements of ${\cal H}_I$.
The space $\h_{II}$ is a Hilbert space under the  inner product:
\beq
\la\vr_1,\vr_2\ra \ \equiv \ \tr (\vr_1^\dagger \ \vr_2)
\label{inner}
\eeq
where the trace is over a complete set of states in ${\cal H}_I$.
Mathematically, a TS, $\vr\in \h_{II}$,  can  always be expended in terms of a
basis $\vr_{\alpha\beta}=|\alpha\ra\la\beta|$ of $\h_{II}$ as
\beq
\vr= \sum C_{\alpha\beta} \vr_{\alpha\beta}
\label{entangled}
\eeq

A general $\vr\in H_{II}$ may not be reducible  to the ``generic form''
(\ref{generic}). A non-generic TS with the ``entangled'' form
(\ref{entangled}) describes situations of a non-complete specification of the
conditions, that is, the final and/or initial conditions correspond to an
entangled state of $\s$ with some other system, say $\s'$, whose degrees of
freedom are traced out.  In
 this case, we have two density matrix $\rho_{in}$ and $\rho_{out}$, rather
then two pure states as conditions. The conditions can be expressed as
$\vr\vr^\dagger|_{t=t_1}  = \rho_{in}(t_1)$ and $\vr^\dagger\vr|_{t=t_2}  =
\rho_{out}(t_2)$. In such circum
stances, the occurrence of an entangled (non-generic) TS is due to the
interaction of  $\s$ and $\s'$ via the measurement device mediator, which is
used to determine the conditions.
Hence the dynamical evolution of the system is not modified (generic TS do not
evolve in time to non-generic or vice versa). The TS of a closed system
satisfies the Liouville  equation:
\beq
i\partial_t\vr = [H , \vr]
\label{evo}
\eeq
In the following we shall study the appearance of  entangled two-states
(\ref{entangled}) in a dynamical way through the  interaction of $\s$ and
$\s'$. To accommodate for this extra  interaction we will need to modify the
Liouville equation (\ref{evo}).

Given by a two-state $\vr(t)$ that corresponds to a pre and post selected
ensemble, we can calculate the quantum mechanical probabilities for the result
of any measurement at time $t$ as follows. Let $A$ be a   Hermitian operator
with a spectral expansion
, $A=\sum aP_a$ in terms of projection operators $P_a=|a\ra\la a|$. Then, the
probability to find  $A=a$ is given by
\beq
\prob(a;t)
= {| \la P_a, \vr(t)\ra|^2
\over
 \sum_{a'}|\la P_{a'}, \vr(t)\ra|^2 }
\label{probII}
 \eeq
Therefore, in analogy with the ordinary expression for probability, the
projection of $\vr$ on $P_a$, $\la P_a, \vr \ra$, can be interpreted as the TS
amplitude. The  absolute square of this amplitude is proportional to the
probability.
In general, this  probability distribution can not be reduced to an expression
in terms of a pure or mixed density matrix.
To see this, notice that Equation (\ref{probII}) can also be written as
\beq
\prob(a;t) = { \la \vr P_a, P_a\vr \ra \over
                  \sum_{a'} \la \vr  P_{a'}, P_{a'}\vr \ra } =
{\tr P_a \rho(a) \over \sum_{a'}\tr P_{a'} \rho(a')}
\label{probII'}
\eeq
where, $\rho(a) \equiv \vr P_a \vr^\dagger$.  Therefore, this probability can
be expressed in terms of a density matrix, only when  $\rho(a)$  is independent
of $a$.

Finally, we note that if the ensemble is only pre (or post) selected, the
ordinary expression for the probability can be obtained as follows.
Assuming that the final (unknown) measurement of some Hermitian operator $\hat
K$ determines one of the eigenstates $\psi_k$, the probability to find $A=a$ is
given by,
\beq
\prob_{I}(a;t)= \sum_{\psi_k} \prob(a) \prob(\psi_k|\psi_{in})
=|\la a|\psi_{in}\ra|^2
\label{ordinary}
\eeq
i.e. by the ordinary expression. In terms of the TS this yields
\beq
\prob_I(a;t)  = {\tr P_a\rho_{in}(t) \over \tr\rho_{in}(t) }
              ={ \la \vr(t), P_a \vr(t)\ra \over \la\vr(t) , \vr(t)\ra }
\label{probI}
\eeq
where $\rho_{in}=\vr\vr^\dagger$. This expression is to be compared with
(\ref{probII'}).
Contrary to the former case of a pre- and post-selection, the latter expression
depends only in the initial condition.

\section{A system with a pre and post selected environment}

Consider a closed system $\s_T$ which is composed of the sub-systems $\s$ and
$\s_e$. Let the part $\s_e$
play the role of an environment $\e$.
The Hamiltonian of the total system is
\beq
H_{tot} = H_s + H_e +  H_{int}= H_0+ H_{int}
\eeq
where $H_s$ and $H_e$ are the "free" Hamiltonians of $\s$ and $\e$,
respectively, and $H_{int}$ is some interaction term.
Given the pre- and post-selected states,  $|\psi_i(t_1)\ra = |s_1\ra \otimes
|e_1\ra$ and $|\psi_f(t_2)\ra = |s_2\ra \otimes |e_2\ra$, the TS in the
Schr\"odinger representation  is
$\vr_{s+e}(t)=U(t-t_1)|\psi_i\ra\la\psi_f|U^\dagger(t-t_2)$, where $U(t_2-t_1)=
\exp(-i\int_{t_1}^{t_2} H_{tot} dt')$.
Limiting out observations only to the subsystem $\s$, we would like to compute
the probabilities for observables of the form $A = A_s \otimes 1_e$, where
$A_s$ operates in  the Hilbert space  $\h_\s$ of $\s$ and $1_e$ is a unit
operator in $\h_\e$.

This probability can be expressed in a simple form by Eq. (\ref{probII}), with
$\vr=\vr_{s+e}(t)$  and $P_a=(P_a)_s \otimes I_e$.
Obviously, since the projection operator acts only in  $\h_\s$,  we can trace
over $\e$ and represent this probability in terms of a reduce TS $\vr_s$ :
\beq
\prob(a_, t | s_2, e_2, s_1, e_1) = {|\la P_a, \vr_s(t)\ra |^2\over
                 \sum_{a'} |\la P_{a'}, \vr_s(t)\ra|^2}
\label{proba}
\eeq
where,
\beq
\vr_s=\vr_s(t; s_2, e_2, s_1, e_1) = {1\over N}\tr_e \vr_{s+e}
\label{reduced}
\eeq
The time independent  normalization, $N=\la
e_2(t_2)|\exp[-iH_e(t_2-t_1)]|e_1(t_1)\ra$,
was chosen  for  later convenience.
At intermediate times $\s$ is completely described in terms of the reduced TS.

Notice that at the boundaries, $t=t_1$ and $t=t_2$, the reduced TS has a simple
generic form:
\beq
\vr_s(t_2) = (\hat U)_w |s_2\ra\la s_1|= |s'\ra\la s_1|
\label{finalts}
\eeq
and
\beq
\vr_s(t_1) = |s_1\ra\la s_2|(\hat U^\dagger)_w = |s_2\ra\la s''|
\label{initialts}
\eeq
where
\beq
(\hat U)_w={\la e_2|\hat U(t_2-t_1)e^{-iH_e(t_2-t_1)}| e_1\ra\over
\la e_2|e^{-iH_e(t_2-t_1)}|e_1\ra}
\eeq
 is the `weak  value' \cite{spin100} of the evolution operator $\hat U$ with
respect to the `free' environment's pre and post-selected states.
Hence, $(\hat U)_w$ is an operator in the Hilbert space $\h_{\s}$.
On the other hand, due to the interaction with the environment, at intermediate
times, $t\in(t_1,t_2)$, the reduced TS is generally a non-reducible
``entangled'' TS:
\beq
\vr_s(t)= \sum C_{s's''}(t)|s'\ra\la s''|.
\label{intermidiatets}
\eeq
This effect of ``decoherence'' and then ``recoherence'' of the reduced
two-state, as expressed in Equations (\ref{finalts}), (\ref{initialts}), and
(\ref{intermidiatets}),
stands in the heart of this paper. The final post selection of the environment
``force's'' the two-state to recohere at the final condition to a generic
two-state.

The effect of post-selecting the environment exists even if the sub-system
$\s$ is not post-selected, i.e. the condition at $t=t_2$ is imposed only on
$\e$.
In this case the probability to find $A=a$ at $t\in(t_2,t_1)$ is given by
\beq
\prob(a, t| r_2, r_1, s_1) = { \sum_{s_2} |\la P_a, \vr_s(s_2)\ra|^2 \over
                                \sum_{s_2,a'}|\la P_{a'}, \vr_s(s_2)\ra|^2 }
\label{parsel}
\eeq

The sum above in over all possible  eigenstates, $\lbrace |s_2\ra\rbrace$, of
an {\it arbitrary} complete set of operator(s) $\hat S$.
This  probability is independent on the choice of $\hat S$.

Although, in this case, there is only one (initial) condition on $\s$, due to
the interaction with the pre- and post-selected environment, the sub-system
$\s$   can not in general be described in terms of a pure or a mixed density
matrix.
Equation (\ref{parsel}) can be rewritten as
\beq
\prob(a, t| r_2, r_1, s_1)= { \tr P_a \rho(a) \over
                                \sum_{a'} P_{a'} \rho(a) }
\label{density}
\eeq
where
\beq
\rho(a) = \sum_{s_2} \vr(s_2) P_a\vr^\dagger(s_2)
\label{ematrix}
\eeq
The object $\rho$ corresponds to a density matrix only if it is independent of
$a$. Intuitively, this happens when the condition at $t=t_2$ on $\e$ does not
``add'' information. Let us examine this question more closely.
When $t\to t_2$, we have $\vr_s\to (\hat U)_w|s_1\ra\la s_1|=|s'\ra \la s_2|$
and by (\ref{ematrix})
\beq
\rho(a,t_2)=\sum_{s_2}|s'\ra\la s_2|a\ra\la a|s_2\ra\la s'| = |s'\ra\la s'|
\eeq
is independent of $a$.
Therefore, near the final condition there is {\it always} an effective  pure
state. The initial state of the open system, $|s_1\ra$ is mapped to  a final
pure state $|s'\ra$ by the ``weak evolution operator''
\beq
\hat U_w |s_1\ra = |s'\ra,
\label{vinitary}
\eeq
Near the initial condition, Eq. (\ref{ematrix}) yields
\beq
\rho(a,t_1)=  ( \la a|(\hat U)_w(\hat U^\dagger)_w |a\ra) |s_1\ra\la
s_1|=C(a)|s_1\ra\la s_1|
\eeq
The effective density matrix is proportional to a pure state, but the
probability at $t=t_1$ depends on an unconventional normalization
\beq
\prob(a,t_1)={\tr P_a\rho(a,t_1)\over\sum_{a'} \rho(a',t_1)}=
{C(a)\over\sum_{a'} C(a')} |\la a|s_1\ra|^2
\eeq
 Unless
$\hat U_w\hat U_w^\dagger=1$, this probability may  depend on the nature of the
final condition on the environment.
For example, if without the post selection we would have $\prob(a)=1$ and
$\prob(b\ne a)=0$, then these probability are not effected by the final post
selection of $\e$.  But the  post selection of $\e$ does generally modify the
probability in intermediat
e cases as
$0<\prob(c)<1$.

At any intermediate times, $t_1<t<t_2$, the effective density matrix
(\ref{ematrix}) will be $a$-dependent, and hence a complete  description in
terms of a  unique  density matrix is not possible.
 It is interesting however, that for a
a limited class of observables, whose nature depends on the coupling with the
environment, we can still construct an effective density matrix. To see this,
let us  choose the (otherwise arbitrary) set $\lbrace |s_2\ra\rbrace$ in
Equation (\ref{parsel}), a
s  eigenvalues of a complete set of an  operators $\hat S_k$ that {\it commute}
with  $H_{int}$.
In this case, for a given  $s_2$ the TS has a generic form :
$\vr(s_2,t) = |s',t\ra\la s_2,t|$.
Therefore, for an operator $A=\sum aP_a$ which is conjugate to one of the
operators $\hat S_k$, we have
$\la s_2|P_a|s_2\ra=constant$. This implies that $\rho(a)$, the effective
density matrix, does not depend on $a$. Hence,  if one measures {\it only} this
limited class of observables, one can use the effective density matrix given by
 $\rho_{den}(t)=\sum_{s_2}\vr\vr^\dagger$.
This density matrix is pure near the conditions at $t=t_1$ and $t=t_2$, but
generally  corresponds to a mixed state at $t_2>t>t_1$.

%

\section{A simple example}

To exemplify these ideas we now consider a solvable model, which was  used to
demonstrate decoherence  \cite{zurek}, of a spin half particle (the system)
coupled to  $N$  spin half particles (the environment).
Setting the free part of the Hamiltonian to zero the interaction part is taken
as
\beq
H_{int} = \sum_{k=1}^N g_k\sigma_z\sigma_z^{(k)}
\eeq
In term of the eigenstates of $\sigma_z$ and $\sigma_z^{(k)}$, the conditions
can be expressed  as
\beq
|\psi_1(t=0)\ra = \biggl(a|\up\ra
+b|\down\ra\biggr)\prod_k\biggl(\alpha_k|\up_k\ra
+\beta_k|\down_k\ra\biggr)=|s_1\ra| e_1\ra
\eeq
and
\beq
|\psi_2(t=T)\ra = \biggl(a'|\up\ra
+b'|\down\ra\biggr)\prod_k\biggl(\alpha'_k|\up_k\ra
+\beta'_k|\down_k\ra\biggr)=|s_2\ra|e_2\ra
\eeq
The reduced TS can be derived according to Eq. (\ref{reduced}), by  tracing
over
the $k=1,..N$ spins. The result is:
$$
\vr_s(t) = {1\over\chi(0)}\biggl\lbrace aa'^*\chi(T)|\up\ra\la\up|
          +bb'^*\chi(-T)|\down\ra\la\down|
$$
\beq
         +ab'^*\chi(2t-T)|\up\ra\la\down|
          +ba'^*\chi(T-2t)|\down\ra\la\up| \biggr\rbrace
\label{exact}
\eeq
where
\beq
\chi(t') = \prod_k\Bigl(\alpha_k\alpha'^*_ke^{ig_kt'}
+\beta_k\beta'^*_ke^{-ig_kt'}\Bigr).
\eeq
At the initial and final conditions, the TS reduces to
\beq
\vr_s(t=0) =|s_1\ra\la s_2|\hat U_w^\dagger(T)
= {1\over\chi(0)}\biggl(a|\up\ra +
b|\down\ra\biggr)\otimes\biggl(a'^*\chi(T)\la\up| +b'^*\chi(-T)\la\down|\biggr)
\eeq
and
\beq
\vr_s(t=T) =\hat U_w(T)|s_1\ra\la s_2|
= {1\over\chi(0)}\biggl(a\chi(T)|\up\ra +
b\chi(-T)|\down\ra\biggr)\otimes\biggl(a'^*\la\up|+
            b'^*\la\down|\biggr)\eeq
where the `weak evolution operator' is
\beq
\hat U_w(t) = {\chi(\sigma_z t)\over \chi(0)}
\eeq

At intermediate times $\vr_s(t)$ can not generally be reduce to a generic TS.

Let us examine the case that only the  $N$ spins (environment) are post
selected.
In this case we need to use equation (\ref{parsel})
and sum over all the final possibilities.
Obviously,  it  is most convenient to sum over final eigenstates of $\sigma_z$.
Hence we have two possible two-states:
\beq
\vr_s(t,\up) =
 {1\over\chi(0)}\biggl(a\chi(T)|\up\ra +
b\chi(T-2t)|\down\ra\biggr)\otimes\la\up|
\eeq
and
\beq
\vr_s(t,\down) ={1\over\chi(0)}\biggl(a\chi(2t-T)|\up\ra + b\chi(-T)|\down\ra
\biggr)\otimes\la\down|
\eeq
The effective density matrix  (\ref{ematrix}) is in this case
$\rho(a)=\vr_s(\up)P_a\vr_s^\dagger(\up)+\vr_s(\down)P_a\vr_s^\dagger(\down)$.
Clearly, if we measure only $\sigma_x$ or $\sigma_y$ this expression reduces to
$\rho_{eff} =
{1\over2}\Bigl(\vr_s(\up)\vr_s^\dagger(\up)+\vr_s(\down)\vr_s^\dagger(\down)
\Bigr)$.
Therefore, for these observables  we have an effective density matrix:
$$
\rho_{eff} =  {1\over2|\chi(0)|^2}\biggl[ \
           |a|^2 \biggl(|\chi(T)|^2 + |\chi(2t-T)|^2\biggr) |\up\ra\la\up|
$$
$$
      +  |b|^2\biggl(|\chi(-T)|^2 + |\chi(T-2t)|^2\biggr)|\down\ra\la\down|
$$
$$
        +  ab^*\biggl(\chi(T)\chi^*(T-2t)
+\chi(2t-T)\chi^*(-T)\biggr)|\up\ra\la\down|
$$
\beq
          +a^*b\biggl(\chi(T-2t)\chi^*(T) +
\chi(-T)\chi(2t-T)\biggr)|\down\ra\la\up| \
\biggl]
\eeq
At the boundaries this expression reduces to
\beq
\rho_{eff}(t=T) = {1\over|\chi(0)|^2}\biggl(a\chi(T)|\up\ra +
b\chi(-T)|\down\ra\biggr)\otimes
             \biggl(a\chi^*(T)\la\up| + b\chi^*(-T)\la\down|\biggr)
 \eeq
and
\beq
\rho_{eff}(t=0) = {1\over2|\chi(0)|^2}\biggl[|\chi(T)|^2 +|\chi(-T)|^2\biggr]
           \biggl(a|\up\ra + b
|\down\ra\biggr)\otimes\biggr(a^*\la\up|+b^*\la\down|\biggr)
\eeq
The initial and final effective density matrix corresponds to a  pure state.
However notice that
the norm of the initial and final pure state is not the same. This reflects the
non-unitarity of the `weak evolution operator'.
Hence, for a limited set of observables, we obtained a description in terms of
a density matrix which initially  decoheres and finally recoheres back to a
pure state.
It is now amusing to note that by fixing the initial and final states of the
$N$ spins to satisfy: $|\alpha_k\alpha'^*_k|^2=|\beta_k\beta'^*_k|^2=1/2$,
(e.g. pre and post selection of  $\sigma_x^{(k)}=1, \ \ \ (k=1,..,N)$), we can
arrange that near the in
itial condition, the state of the system is described for {\it any} observable
by a pure state. In this case the  system in intermediate time is
(effectively), for some observables, in a mixed state, while for  other
observables, even a mixed state not
exits. The system always `recoheres' back to a pure state.

\section{Reduced two-state dynamics}

The  two-state of a closed system satisfies a Liouville
Equation. By focusing on a subsystem, and tracing over the environment's
degrees of freedom we will also modify the equation of motion of the the
reduced two-state. Some additional terms are now necessary to  accommodate for
the effect of the `external' e
nvironment.
This problem is reminiscent to the well studied issue of environment induced
decoherence.
 There is however a significant difference between the two problems. As we have
seen, when the conditions correspond to pure states, the {\it exact} solution
for the TS must be of generic (direct product) form, {\it both}, initially at
$t=t_1$ and  finall
y at $t=t_2$. Therefore, the resulting dynamical equation must have the
non-trivial property that given any two conditions for $\s$, it evolves an
initially generic  TS to an ``entangled TS'' at intermediate times, and back to
a generic TS at at the final
 condition.  Such a `fine
tuning' requires cushion when approximations are used to derive the corrections
to the Liouville Equation.
For example, in deriving the equation of motion to the reduced density matrix,
it is usually assumed that one can use the `non-reversible'  approximation that
the density matrix can be factorized to a product of two density matrix of
form:  $\rho_{density
} = \rho_s\times\rho_e$. This simplifies considerably the computations.
However, in our case such an approximation is invalidated since the TS can not
be factorized in such a way at any time.
In fact a naive usage of such a factorization
leads to an equation of motion with no solutions for the two boundary condition
problem.

In the following we shall derive  perturbatively the  modified Liouville
Equation. Therefore we expect our solution to be valid only in the weak
coupling regime  $\lambda T < 1$, where $\lambda$ is the coupling constant (
$H_{int}=\lambda H_I$), and $T=t_2-t_1$.
For simplicity we shall assume a time independent Hamiltonian and that
$H_{int}$ is an analytic function.
In the following, it will be most convenient to use the interaction
representation.
Setting  $t_1=0$ and $t_2=T$ we define the TS in the interaction representation
as
\beq
\vr_{int}(t) = e^{iH_0t} \vr(t) e^{-iH_0t}
\eeq
The equation of motion of the closed system is
\beq
\partial_t \vr_{int} = -i \Bigl[ [H_{int}]_I, \vr_{int} \Bigr]
\eeq
where $[\o]_I\equiv e^{iH_0t}[\o]e^{-iH_0t}$.

Now define $\vr_0 = |\psi_1(0)\ra\la\psi_2(T)|
\exp(-iH_0T)=\vr_{s0}\otimes\vr_{e0}$, which is the free ($H_{int}=0$)
two-state at $t=0$. In terms of $\vr_0$ we have
\beq
\vr_{int}(t) = \biggl[ e^{-iHt}e^{+iH_0t}\biggr]_I \vr_0
\biggl[e^{-iH_0(t-T)}e^{+iH(t-T)}\biggr]_I
\eeq
For simplicity let us assume that  $[H_0, H_{int}]=0$, hence
\beq
\vr_{int}(t) = e^{-i[H_{int}]_I t} \vr_0 e^{+i[H_{int}]_I (t-T)}
\label{vrint}
\eeq
Although the exact solution $\vr_{int}$ can not be factorized, we can use
(\ref{vrint}) to expend it in powers of  $\vr_0=\vr_{s0}\times\vr_{e0}$.
Putting from now on $\vr = \vr_{int}$ and $\lambda H_I= [H_{int}]_I$, we have:
\beq
\vr(t) = \vr_0 -i\lambda t H_I  \vr_0 - i\lambda(T-t)\vr_0 H_I +O(\lambda^2)
\label{expend}
\eeq
The free TS $\vr_0$ is factorizable, and we can now trace over $\e$. Therefore,
\beq
\vr_s(t) \equiv {\tr_e\vr\over\tr_e\vr_{e0}} = \vr_{s0} - i\lambda t(H_I)_w
\vr_{s0}
                    -i\lambda(T-t)\vr_{s0}(H_I)_w +O(\lambda^2)
\label{bexpend}
\eeq
where $(...)_w$ stands for the weak value with respect to free environment's
two-state,  and is defined by $\o_w=\tr\o\vr_{e0}/\tr\vr_{e0}$.
The last expression can be also inverted to
\beq
\vr_{s0} = \vr_{s}(t) + i\lambda t (H_I)_w \vr_{s}(t)
                     + i\lambda (T-t)\vr_{s}(t)(H_I)_w +O(\lambda^2)
\label{reexpend}
\eeq

Substituting  (\ref{expend}) into  the Liouville equation  and tracing over
 $\e$ yields
\beq
\partial_t \vr_s(t) = -i\lambda[(H_I)_w, \vr_{s0}]
                      - \lambda^2\biggl[\Bigl(H_I , t H_I  \vr_{s0}
                   +(T-t)\vr_{s0} H_I \Bigr)_w \biggr] +O(\lambda^3)
\eeq
Finally, we can use (\ref{reexpend}) to reexpress the last equation in terms of
$\vr(t)$. We  get
\beq
\partial_t \vr_s(t) = -i\lambda[(H_I)_w, \vr_{s}(t)]
\eeq
\beq
                      -\lambda^2 \biggl[\Bigl(H_I , t H_I  \vr_{s}(t)
                   +(T-t)\vr_{s}(t) H_I \Bigr)_w \biggr] +
     \biggl[(H_I)_w , \Bigr(t H_I \vr_{s}(t)
                   +(T-t)\vr_{s}(t)  H_I \Bigr)_w \biggl] +O(\lambda^3)
\eeq

Let us consider some  examples.
For a generic interaction of the  form:
\beq
H_I = \lambda Q_i L_i
\eeq
where the $Q_i$'s are some system variables and $L_i$ reservoir variables,
we get in the free case ($H_s=H_e=0$):
\beq
\partial_t\vr_s(t) = - i\lambda (L_i)_w [Q_i, \vr_s]
                   -\lambda^2 \Delta_{ij}[Q_i, \ tQ_j \vr_s + (T-t)\vr_s Q_j] +
O(\lambda^3)
\label{QLint}
\eeq
where
\beq
\Delta_{ij} =   (L_iL_j)_w - (L_i)_w(L_j)_w
\eeq

Typically, the first order is the Liouville equation with a ``weak"
Hamiltonian''.
The second order corrections, are proportional to the ``weak uncertainty''
$\Delta_{ij}$. Higher order may be easily computed, but become very cumbersome.
It is straightforward to rewrite (\ref{QLint}) to the case that $L_i$ and $Q_i$
are not constants of motion, or to any other polynomial interaction.

Simplifying the interaction even further, we set $Q_1=\sigma_z$, $L_1=L_z\equiv
L$ and
$L_i=Q_i=0$ for $i\ne1$.
This corresponds to a spin half subsystem which interacts with
the $z$ component of the angular momentum of the environment.
Equation (\ref{QLint}) reduces to
\beq
\partial_t \vr_s = -i\lambda L_w [\sigma_z, \vr_s]
                   -2\lambda^2\Delta L_w (2t - T)(\vr_s -
\sigma_z\vr_s\sigma_z ) + O(\lambda^3)
\label{SLint}
\eeq
where $\Delta L_w =  (L^2)_w - (L_w)^2$.

 We can easily verify that for every two initial and final conditions for $\s$,
 there exists an appropriate solution.
It is only the second order term that can induce transition from generic to
non-generic (entangled) two-state. In terms of the notation
$\vr_{\up\down} =|\up\ra\la\down|$, etc, the general solution of Eq.
(\ref{SLint}) is
\beq
\vr_{\up\up}(t) = \vr_{\up\up0}, \ \ \ \vr_{\down\down}(t)=\vr_{\down\down0}
\eeq
\beq
\vr_{\up\down}(t) = \exp\biggl[ -i2\lambda L_wt -4\lambda^2\Delta L_w(t^2 - Tt)
\biggr] \vr_{\up\down0}
\eeq
\beq
\vr_{\down\up}(t) = \exp\biggl[ +i2\lambda L_wt -4\lambda^2\Delta L_w(t^2 - Tt)
\biggr] \vr_{\down\up0}
\eeq
Clearly, due to the factor $t^2-Tt$, the second order contributions vanishes on
at the  conditions.
By substituting $\lambda L = \sum g_k\sigma_z^k$, it can be verified that this
solution agrees up to corrections of order $O(\lambda^3)$ with the exact
solution given by equation (\ref{exact}).

Due to the continues interaction with each of the spins in the latter problem,
the validity of Equation (\ref{SLint}) is limited by the constraint
$T<1/\lambda$. We shall now compare this system to the other extreme case, in
which the subsystem interacts
with each of the particles of the environment  separately, and only for a very
short time  $\Delta t= \tau$, such that $\tau\lambda<<1$. In this way the weak
coupling condition is satisfied, and our modified Liouville Eq. can be applied
also for long tim
es.
Let the environment be composed of $N$ non-interacting particles.
The interaction Hamiltonian for this case  is given by \cite{davydov}
\beq
H_I = \lambda\sum_{n=1}^N f_n(t) H_n
\eeq
where $f_n(t) = \theta(t-n\tau) - \theta(t-(n+1)\tau)$ with $\theta(t)$ as the
step function is nonzero only for $t\in(n\tau, (n+1)\tau)$.  $H_n$ is the
interaction of $\s$ with the $n$th particle. Let us further assume that $H_n$
can be regarded as (or i
s)  constant during the interaction times $\tau$.
For $H_n=\sigma_zL_{nz}=\sigma L_n$ we get
$$
\partial_t \vr_s = -i\lambda \sum_n f_n(t) L_{nw}[\sigma , \vr_s]
             -\lambda^2\sum_{n} f_n\Delta_{nn} (2t-(2n+1)\tau)
                  (\vr_s - \sigma\vr_s\sigma)
$$
\beq
         -\lambda^2\sum_{n,m=1}^{m=n-1}f_n\Delta_{nm}(n\tau)[\sigma,
\sigma\vr_s]
            -\lambda^2\sum_{n,m=n+1}^{m=N}f_n\Delta_{nm}(N-n-1)\tau[\sigma,
\vr_s\sigma]
\label{reseq}
\eeq
where $\Delta_{nm} = (L_nL_m)_w - (L_n)_w(L_n)_w$.
If the  initial and final states of the environment are given by a product
state,  $\prod_k\otimes |e_k\ra$ of the $N$ particles, there are no
correlations between the weak values different particles in the reservoir and
$\Delta_{nm} = ((L_n^2)_w - (L_{nw
})^2 )\delta_{nm}$.
Therefore, in this case the two last terms on the right hand side of equation
(\ref{reseq}) vanish.
Integrating (\ref{reseq}) we see that after each ``step'', when the
interaction with the $n$'th particle in the environment is completed, the
accumulated contribution of the second term drops to zero. The TS remains
`pure' up to fluctuations of order $O(
\lambda^2\tau^2)$.
In this sense, we can say that the post-selection of the environment prevents
decoherence of the subsystem.

t a time scale $T$.

\vspace{.5in}
\noindent
{\bf Acknowledgment}

I would like to thank Bill Unruh for helpful discussions.

\vfill\eject

\end{document}